# Optical Spectroscopic Imaging for Early Detection of Salinity Stress in Plants: A Review

[1]Ramji Gupta; [2]Snehprabha Gujrathi; [3]Swati Sharma; [4]Saurav Bharadwaj

[1234]Parul Institute of Engineering and Technology, Parul University, Vadodara, Gujarat, India

[4]Corresponding Author: Saurav Bharadwaj, saurav.bharadwaj33162@paruluniversity.ac.in

ORCID: 0000-0003-1882-7287

**Abstract:** Salinity stress poses a significant challenge to global agriculture, necessitating efficient and scalable approaches for early detection and management. This review examines advanced optical spectroscopic imaging techniques, including fluorescence imaging, hyperspectral imaging, and thermal imaging, for their ability to detect physiological and biochemical responses in plants subjected to saline conditions. By focusing exclusively on imaging-based optical methods and excluding traditional spectroscopic techniques, the review emphasises non-invasive tools capable of delivering high sensitivity and specificity. These technologies enable real-time, high-throughput analysis, making them particularly well-suited for large-scale plant screening and breeding programs targeting salt tolerance. The integration of ground-based optical systems with aerial platforms such as unmanned aerial vehicles and satellites significantly expands spatial and temporal monitoring capabilities.

## 1 Introduction

Salinity stress is a widespread abiotic factor that significantly limits global agricultural productivity by inducing physiological and morphological changes in plants. Recent climate change projections predict increased soil salinisation, particularly in arid and semi-arid regions. Rising global temperatures, altered precipitation patterns, and the intensification of evapotranspiration processes contribute to the accumulation of soluble salts in the root zone, exacerbating salinity stress. Sea-level rise and increased frequency of extreme weather events are expected to accelerate saltwater intrusion into coastal agricultural lands. An increase in soil salinity diminishes the osmotic potential of the soil solution, impeding water absorption by roots and inducing osmotic stress, which consequently decreases cell turgor pressure [1]. In response, plants close their stomata to conserve water; however, this defence mechanism also restricts carbon dioxide uptake, thereby diminishing photosynthetic efficiency [2]. Excessive accumulation of sodium ($Na^+$) and chloride ($Cl^-$) ions disrupts ionic equilibrium, impeding the absorption of vital nutrients such as potassium ($K^+$), calcium ($Ca^{2+}$), and magnesium ($Mg^{2+}$), which leads to metabolic dysfunction and cellular damage. To alleviate ion toxicity, plants engage energy-demanding systems, such as ion transporters and vacuolar sequestration of $Na^+$, which may prove inadequate under extended stress conditions [3]. Photosynthesis is impeded by direct ionic toxicity and oxidative damage to chloroplast structures, frequently resulting in chlorophyll loss and observable leaf chlorosis. Stress circumstances also lead to the excessive generation of reactive oxygen species (ROS), such as superoxide anions ($O_2^-$), hydrogen peroxide ($H_2O_2$), and hydroxyl radicals (•OH), which can result in oxidative damage to proteins, lipids, and nucleic acids [4]. In response, plants augment their antioxidant defence mechanisms by upregulating enzymes such as superoxide dismutase (SOD), catalase (CAT), and different peroxidases to preserve redox equilibrium [5].

Conventional approaches for identifying salinity stress primarily depend on physiological, biochemical, and morphological signs, each of which, however helpful, has intrinsic limits. Morphological indicators, such as chlorosis, necrosis, and wilting, are easily noticeable but typically appear only after significant damage has occurred, thereby constraining their utility for early identification [6], [7]. Quantitative evaluations of growth metrics, including shoot and root length, leaf area, and biomass, offer comprehensive insights into the effects of stress; however, they are generally endpoint measurements that require destructive sampling [8]. Physiological parameters, including relative water content, leaf water potential, stomatal conductance, transpiration rate, and photosynthetic activity, provide critical insights into plant water relations and metabolic status under stressful circumstances [9]. These evaluations often require specialised equipment and are not readily scalable for high-throughput screening of large populations. Biochemical assays, such as lipid peroxidation measures and antioxidant enzyme activity assessments, provide mechanistic insights into oxidative and ionic stress responses [2], [10]. Although these approaches exhibit great accuracy, they are time-consuming, require substantial sample preparation, and are typically destructive, hence imposing considerable constraints on longitudinal or high-throughput applications [11].

Progress in optical sensing technology offers potential, non-invasive methods for identifying salt stress by real-time monitoring of plant physiological conditions [12]. Methods such as hyperspectral imaging, fluorescence imaging, and thermal imaging provide ongoing evaluation of plant responses across time without necessitating physical sampling [13]. These approaches can identify modest, early physiological alterations—such as changes in pigment composition, water content, and photosynthetic efficiency—prior to the appearance of apparent symptoms, facilitating prompt and informed management decisions [14]. Optical sensing methodologies are intrinsically high-throughput, enabling the swift evaluation of extensive plant populations with reduced labour and operating expenses [15], [16]. Their interoperability with automated data collecting and computational analysis systems renders them especially appropriate for incorporation into plant breeding programs and precision agriculture platforms. Unlike conventional biochemical tests, optical approaches provide a more scalable and efficient approach for early stress detection and phenotypic screening in salinity stress research [17], [18].

Effective mitigation of salinity stress requires an integrated strategy that combines physiological, molecular, and agronomic approaches to address the complex nature of this stress. (1) Breeding and biotechnological interventions aim to develop salt-tolerant cultivars characterized by traits such as enhanced $Na^+$ exclusion, maintenance of $K^+/Na^+$ balance, and upregulation of stress-responsive genes [19]. Genes encoding ion transporters, such as HKT1 and SOS1 (salt overly sensitive), as well as those involved in the biosynthesis of osmoprotectants, play pivotal roles in conferring tolerance. CRISPR/Cas9-mediated genome editing has further accelerated the development of such genotypes by enabling precise modification of stress-related loci [20]. (2) Complementing genetic approaches, exogenous application of plant growth regulators—including abscisic acid, salicylic acid, and brassinosteroids—can alleviate salinity-induced damage by enhancing antioxidant defences, stabilizing cellular membranes, and improving water use efficiency [21]. (3) Biological interventions involving plant growth-promoting rhizobacteria and arbuscular mycorrhizal fungi contribute to induced systemic tolerance, bolster nutrient uptake, and promote favourable root architecture [22]. (4) Agronomic practices such as drip irrigation help control soil moisture and salt accumulation, while soil amendments like gypsum and organic matter improve soil structure and ion exchange capacity [7]. Mulching further diminishes surface evaporation and salt accumulation. In conjunction with early diagnostic techniques, these strategies provide prompt intervention and sustainable management of salt stress, enhancing crop resilience and yield.

This study is significant for its thorough assessment of sophisticated optical spectroscopic imaging methods for the early identification and ongoing observation of salinity stress in plants, a matter of growing concern due to climate change and soil degradation. The article critically evaluates techniques, including fluorescence imaging, hyperspectral imaging, and thermal imaging, emphasising their accuracy in detecting early physiological and biochemical reactions to salt stress. The techniques developed over the last twenty years offer considerable benefits compared to conventional laboratory methods, including high-throughput data collection, non-destructive analysis, and real-time analysis, making them essential for extensive screening in plant phenotyping and breeding initiatives. Moreover, their incorporation with drone and satellite-based remote sensing systems enhances their utility for field-level monitoring, facilitating spatial mapping of salinity-impacted areas. The integration of proximal and remote sensing technology enables precision agriculture by providing timely, site-specific management techniques, thereby enhancing crop resilience and promoting sustainable farming practices.

## 2 Biological Interpretations

### 2.1 Physiological and Morphological Effects

Salinity stress presents a complex challenge to plant systems by inducing a series of osmotic and ionic disruptions that together hinder physiological and biochemical functions. The initial osmotic stress phase arises from a decrease in soil water potential, which limits water uptake, causes reduced cell turgor, stomatal closure, and inhibited cell expansion [23]. Prolonged exposure transitions into the ionic phase, characterized by the accumulation of $Na^+$ and $Cl^-$ ions, which disrupt essential nutrient uptake, compromise membrane stability, and inhibit enzymatic and metabolic functions. These ionic disruptions further exacerbate oxidative stress by overproducing ROS, resulting in extensive damage to lipids, proteins, and nucleic acids [24]. Photosynthetic efficiency deteriorates due to chlorophyll degradation, stomatal limitations, and impaired activity of photosystem II [25]. Plants exhibit stunted germination, reduced seedling vigour, chlorosis, necrosis, leaf wilting, and damaged root architecture, as shown in Figure 1(a) and 1(b). Reproductive development is also highly sensitive, with marked declines in pollen viability, fertilization rates, and overall seed and fruit formation [26]. At the cellular level, osmotic imbalance and ionic toxicity result in structural disintegration, including plasmolysis, vacuolar collapse, and cytoplasmic shrinkage, all contributing to significant reductions in plant growth under saline conditions [6], [27].

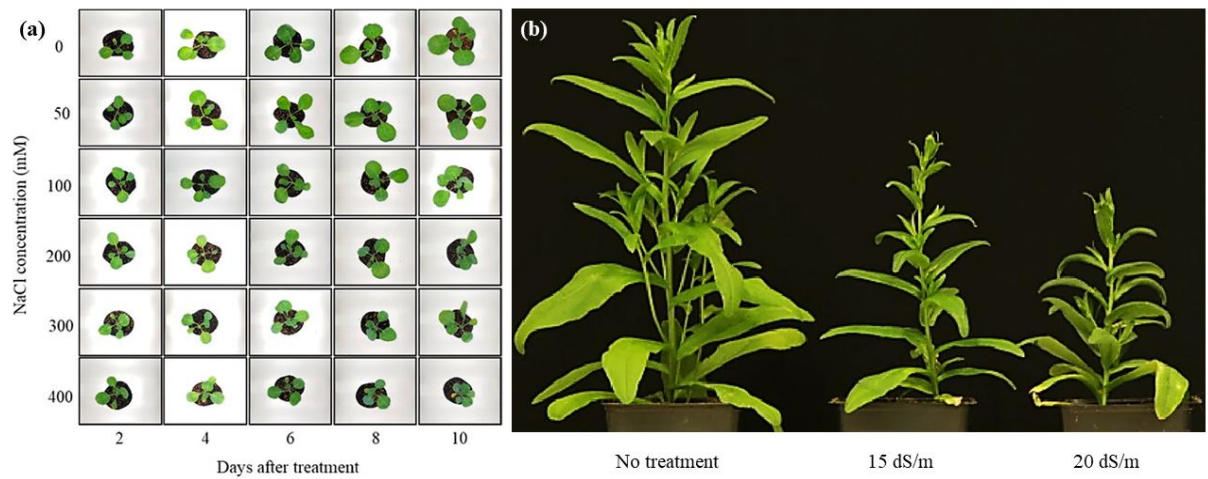

Figure 1 (a) Time-dependent reduction in total leaf area of Brussels sprouts subjected to salt treatment, illustrating the impact of salinity stress over time [28]. (b) Growth comparison of plants after 28 days, showing effects of salt treatments at 15 dS/m and 20 dS/m applied 20 days post-sowing, alongside control plants grown without NaCl [29].

## 2.3 Adaptation Mechanisms

Plants respond to salinity stress through a coordinated array of physiological, biochemical, molecular, and structural adaptations. A key mechanism is osmotic adjustment, involving the accumulation of compatible solutes like proline, glycine betaine, and sugars to retain water, protect cellular structures, and maintain redox balance [30], [31]. To counter ion toxicity, plants regulate $Na^+$ and $Cl^-$ uptake, sequester excess ions into vacuoles via $Na^+/H^+$ antiporters (e.g., NHX1), and maintain a favourable cytosolic $K^+/Na^+$ ratio. The SOS pathway—comprising SOS3, SOS2, and SOS1—plays a central role in $Na^+$ extrusion, activated through calcium signalling and possibly GIPC-mediated $Na^+$ sensing. The ROS, generated by NADPH oxidases (RbohD/F), function as secondary messengers in salt stress signalling, promoting $Ca^{2+}$ influx and activating stress-responsive transcription factors via MAPK cascades [32]. Hormonal pathways, especially abscisic acid (ABA) signalling through SnRK2 kinases and PYR/PYL-PP2C modules, further integrate into this regulatory network to modulate gene expression and stress adaptation [33].

Salinity also triggers cell wall surveillance mechanisms via the LRX-RALF-FERONIA complex, which detects structural damage and initiates repair through $Ca^{2+}$ signalling and pH regulation. Vacuolar transporters (NHXs, CAX1, TPK1) and $H^+$-ATPases maintain intracellular ion and pH homeostasis under stress conditions [34].

Morphologically, plants exhibit adaptations such as succulence, the formation of salt glands or bladders, altered stomatal density, increased root-to-shoot ratios, and reduced shoot growth, which limit salt accumulation and water loss [35]. These responses collectively enhance salt tolerance by maintaining cellular integrity, optimising water use, and minimising ion toxicity, as shown in Figure 2.

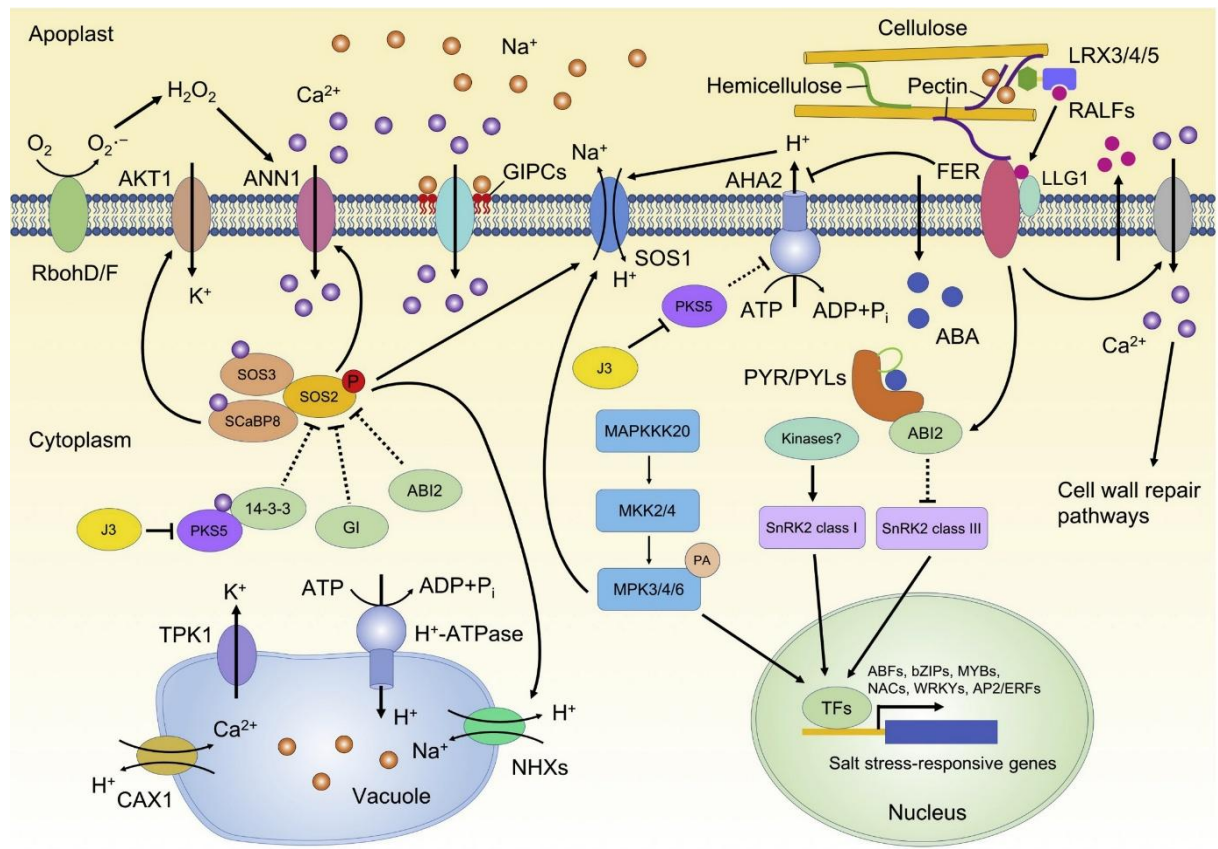

Figure 2 The plant salt stress signalling network integrates multiple pathways to maintain cellular homeostasis and activate adaptive responses. Central to this network is the SOS pathway (SOS3/SCaBP8–SOS2–SOS1), which regulates ion homeostasis by extruding excess $Na^+$ in response to salt-triggered cytosolic $Ca^{2+}$ signals. Negative regulators, such as 14-3-3 proteins, GI, and ABI2, inhibit SOS2 kinase activity; however, this inhibition is relieved through a $Ca^{2+}$-dependent interaction with PKS5. Glycosyl inositol phosphorylceramides (GIPCs) function as putative $Na^+$ sensors, initiating $Ca^{2+}$ influx essential for SOS pathway activation. The ROS produced by NADPH oxidases RbohD/F stimulate ANN1-mediated $Ca^{2+}$ signalling. The AKT1 channel regulates $K^+$ uptake to maintain ionic balance. At the same time, MAP kinase cascades and SnRK2 kinases, operating through both ABA-dependent and independent routes, relay salt stress signals to transcription factors including ABFs, bZIPs, MYBs, NACs, WRKYs, and AP2/ERFs, thereby modulating stress-responsive gene expression. The apoplastic LRX–

RALF–FER complex senses cell wall disturbances, triggering Ca²⁺ signalling and activating cell wall repair mechanisms; FER also controls the plasma membrane H⁺-ATPase AHA2 to adjust apoplastic pH. Vacuolar transporters, such as NHXs, CAX1, TPK1, and H-ATPase, contribute to intracellular ion sequestration and pH homeostasis under salinity stress. Dashed connections in the network represent the release of negative regulation under salt stress, allowing for the full activation of these signalling components [36].

## 3 Optical Spectroscopic Imaging Techniques

Optical screening techniques are non-destructive, high-throughput approaches widely employed to monitor and evaluate plant responses to salinity stress by leveraging the interaction of light with plant tissues to detect physiological and biochemical alterations [16], [37]. Chlorophyll fluorescence imaging, for instance, captures the re-emission of absorbed light by chlorophyll molecules during photosynthesis, primarily from photosystem II, offering insights into photosynthetic efficiency and stress-induced impairments [38], [39]. Hyperspectral imaging provides detailed spectral reflectance data across a broad wavelength range (400–2500 nm), enabling precise assessment of pigment composition and early detection of stress signatures [40], [41]. Thermal infrared (IR) imaging quantifies variations in canopy or leaf surface temperature, which serve as indirect indicators of transpiration rates and stomatal conductance, both of which are commonly affected by salinity stress [42]. Optical techniques offer real-time tools for high-resolution phenotyping and stress diagnostics in plants [16], [43].

### 3.1 Fluorescence Imaging

Fluorescence imaging is a highly sensitive and widely employed technique in biological research for visualising and quantifying the distribution of fluorescent molecules within a sample. This method relies on the fundamental principle that specific molecules, either intrinsically fluorescent or labelled with fluorophores such as synthetic dyes, can absorb light at a defined excitation wavelength and re-emit it at a longer emission wavelength [44]. When stimulated by a light source, usually within the ultraviolet or visible spectrum, the fluorophore temporarily transitions to an excited electronic state before reverting to its ground state, generating fluorescence throughout this process [45]. The spectral disparity between excitation and emission wavelengths, referred to as the Stokes shift, enables the selective detection of the emitted signal by eliminating the excitation light [46]. Specialised detectors, such as charge-coupled devices or photomultiplier tubes, capture the resulting fluorescence, generating high-resolution images of biological specimens [38].

A typical fluorescence imaging configuration includes a light source—usually a laser or a high-intensity xenon or mercury arc lamp—to excite the fluorophores. The emitted fluorescence passes through an emission filter that blocks leftover excitation light, ensuring precise signal detection [47]. Dichroic mirrors and bandpass filters are systematically incorporated into the optical channel to selectively transmit specific wavelengths selectively, thereby improving the signal-to-noise ratio and reducing background fluorescence [48]. Fluorescence imaging systems may utilise either wide-field or confocal designs, depending on the application [49]. Wide-field systems uniformly illuminate and capture the entire sample, facilitating rapid image acquisition, while confocal systems employ point illumination and spatial pinholes to eliminate out-of-focus light, providing enhanced resolution images of biological structures[50].

The pre-processing of fluorescence pictures is essential for ensuring precise and dependable data analysis. The procedure generally begins with the evaluation of image acquisition quality, where photos are scrutinized for issues such as saturation, focus, or inconsistent illumination. The subsequent stage is background subtraction, which entails eliminating autofluorescence or nonspecific signals to improve the contrast between the signal of interest and the background. This can be accomplished by techniques such as rolling ball algorithms or local thresholding [51]. Denoising is then employed to mitigate random noise generated by the imaging equipment, frequently utilizing Gaussian, median, or non-local means filters [52]. After noise reduction, image registration may be necessary, especially in multi-channel or time-lapse imaging, to align images spatially and correct for drift or movement [53]. Flat-field correction is also important for addressing uneven illumination across the field of view by normalizing image intensities based on a reference flat-field image [54]. In some cases, intensity normalization is performed to ensure consistent brightness levels across different images or experiments [45]. Channel separation and spectral unmixing are performed in multi-colour fluorescence imaging to correct for bleed-through or crosstalk between channels. These pre-processing steps collectively enhance the quality of fluorescence images, enabling more accurate downstream analysis [38], [55].

**3.2 Hyperspectral Imaging**

Hyperspectral imaging is an advanced optical sensing technology that captures detailed spectral information across a broad range of the electromagnetic spectrum, extending beyond the capabilities of conventional RGB imaging [56]. Unlike traditional imaging, which records data in just three broad bands—red, green, and blue—hyperspectral imaging divides the spectrum into numerous narrow, contiguous spectral bands, typically spanning the visible (400–700 nm), near-IR (700–1000 nm), and often extending into the short-wave IR (1000–2500 nm) or beyond [57]. This acceptable spectral resolution enables each pixel in a hyperspectral image to contain a full reflectance spectrum, delivering rich spatial and spectral information simultaneously [58]. Central to hyperspectral imaging is the formation of a three-dimensional dataset, known as a hypercube, which is generated by integrating a spectrometer with an imaging sensor, as shown in Figure 3. The hypercube comprises two spatial dimensions and one spectral dimension, where each voxel represents the intensity of reflected or emitted light at a specific wavelength for a specific location [59], [60].

Hyperspectral imaging systems utilize three principal ways of data acquisition: whiskbroom (point scanning), pushbroom (line scanning), and snapshot (area scanning). Whiskbroom systems employ a single-point detector that scans the scene in both spatial dimensions, providing excellent spatial resolution, albeit with slower data acquisition rates [61]. Pushbroom scanners utilize a linear array of detectors to collect a single line of spatial data at a time, offering an effective balance between spatial and spectral fidelity, along with increased acquisition speed and fewer moving parts. Snapshot methods capture comprehensive 2D spatial data across all spectral bands in a single exposure, facilitating real-time [62]. Pushbroom technology is predominantly utilized in remote sensing applications within agriculture, where the collection of efficient, high-resolution spectral data is essential [63].

The pre-processing of hyperspectral pictures is essential for guaranteeing the quality of subsequent analysis [64]. The procedure generally commences with radiometric calibration, wherein unprocessed sensor data is transformed into reflectance values by rectifying sensor-specific aberrations and dark current [65]. The subsequent step is spectral calibration, which ensures the accurate alignment of each spectral band with its corresponding wavelength, thereby rectifying any spectral misalignments [66]. Subsequently, geometric correction is implemented to spatially align the hyperspectral image with ground-truth maps or other image data, addressing errors caused by sensor movement or terrain fluctuations [67]. Noise removal is a crucial process, as hyperspectral images frequently exhibit noisy or redundant bands; methodologies such as Minimum Noise Fraction processing or Principal Component Analysis are utilised to reduce dimensionality and improve signal integrity [68]. Atmospheric correction is performed to eliminate the impacts of air scattering and absorption, hence deriving surface reflectance from at-sensor radiance. Bad band removal is conducted to eliminate spectral bands that are significantly compromised by noise or ambient water absorption, such as those between 1400 nm and 1900 nm. Image normalisation guarantees consistency in pixel intensity throughout the image, facilitating improved comparisons across different scenes or temporal intervals. These pre-processing procedures guarantee that the hyperspectral data is calibrated for significant analysis [61].

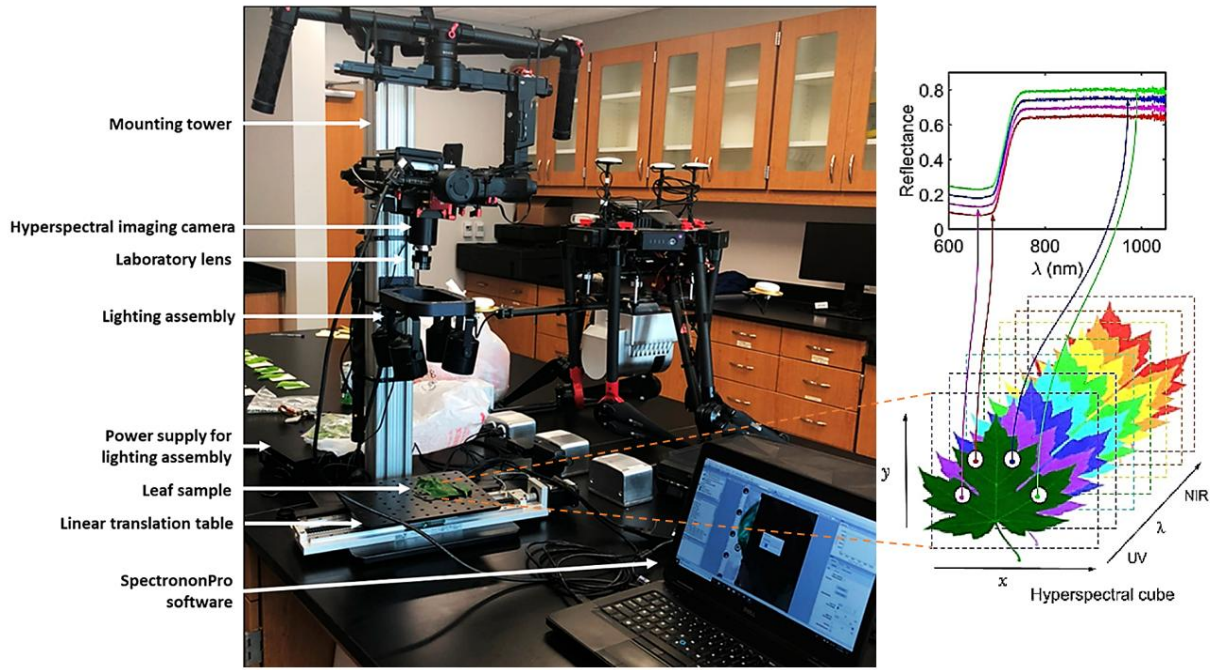

Figure 3 Hyperspectral imaging system and indoor experimental setup used for capturing hyperspectral images of a plant leaf [16], [69].

### 3.3 Thermal Imaging

Thermal imaging, also known as IR thermography, is a method that detects and visualises the infrared radiation emitted by objects, converting it into a visible representation termed a thermogram—all objects with a temperature exceeding absolute zero produce infrared radiation proportional to their thermal energy [42]. The properties of this emission—its intensity and spectrum distribution—are governed by Planck's law and are affected by the temperature and emissivity of the object [70]. Thermal imaging systems are generally engineered to identify radiation in the mid-wave infrared (MWIR, 3–5 µm) and long-wave infrared (LWIR, 8–14 µm) spectral ranges, where thermal emissions from objects at ambient temperatures are most significant [71].

A thermal imaging camera consists of numerous critical components, including an infrared sensor, an optical system, and signal processing circuits [72]. The infrared sensor, often a focal plane array detector composed of materials like vanadium oxide or indium antimonide, detects incoming heat radiation and converts it into electronic impulses [73]. The signals are subsequently digitised and processed to generate a two-dimensional image, where the intensity of each pixel represents the corresponding temperature distribution of the recorded scene. Advanced thermal imaging systems typically incorporate real-time image processing, temperature adjustment mechanisms, and calibration algorithms to enhance measurement spatial resolution and image quality [74], [75].

The pre-processing of thermal images is crucial for enhancing their quality and ensuring accurate analysis [76]. The process typically begins with noise reduction, as thermal cameras often generate sensor-related noise that can impact temperature measurements [77]. Gaussian filtering, median filtering, and non-local means denoising are frequently utilized to enhance image smoothness while maintaining critical edges. Image normalization is applied to scale pixel intensity values to a standard range, making it easier to compare images captured under different conditions [78]. This is followed by contrast enhancement, such as histogram equalisation or contrast-limited adaptive histogram equalisation, which improves the visibility of features by adjusting the distribution of pixel intensities. In cases where thermal images are used for object detection or segmentation, background subtraction and thresholding may be applied to isolate regions of interest [79]. Geometric corrections are sometimes necessary to align images captured at different angles or times, especially in time-series analysis. Calibration is performed using reference blackbody sources or known temperature points to ensure that the temperature data in the image corresponds accurately to real-world values [80], [81].

## 4 Practical Applications in Salinity Studies

### 4.1 Fluorescence Imaging Applications

A multicolour fluorescence imaging system was developed to facilitate early, non-invasive detection of salt stress in plants. The system integrated a 365 nm excitation LED panel ($48 \times 3W$), a monochrome charge-coupled device camera with a resolution of $1024 \times 1280$ pixels, a 12 mm focal length lens, and four band-pass filters centred at 440 nm, 520 nm, 690 nm, and 740 nm (half-bandwidth of 15 nm). Four mirrors were mounted on the inner walls of a custom-designed dark imaging box to achieve uniform illumination throughout the space. Plant specimens were positioned 25 cm from the camera on a stationary platform, and consecutive multicolour fluorescence images were obtained with four filters to capture unique spectral signatures. A dedicated image acquisition program, developed in C++, facilitated real-time image capture. Multiple fluorescence parameters were extracted and analysed via principal component analysis to reduce dimensionality and identify the most discriminative features. These features were used to train a support vector machine classifier, which achieved 92% classification accuracy at 5 days post salt treatment and 98% by day 9, demonstrating the effectiveness for early-stage salt stress detection in plants, as shown in Figure 4(a) [82].

A separate study developed a high-throughput procedure to assess salinity tolerance in two Indica rice cultivars (IR64 and Fatmawati) under different levels of salt stress. Visible RGB and fluorescence imaging were utilised to measure shoot growth and senescence under moderate and high salt conditions. Image acquisition was captured in side and top views along with fluorescence data. Image processing and analysis were conducted, enabling precise quantification of biomass and senescence metrics. Results indicated that shoot area growth declined while senescence increased with higher NaCl concentrations, particularly after 20 days, corresponding to elevated shoot $Na^+$ accumulation. Differences in senescent area between IR64 and Fatmawati under high salt stress suggest cultivar-specific tissue tolerance mechanisms. This image-based phenotyping approach effectively distinguished between the osmotic and ionic phases of salt stress in rice, as shown in Figure 4(b) [83].

A separate investigation employed chlorophyll fluorescence imaging, combined with deep learning, to rapidly and accurately identify salt stress in soybeans. Six classical convolutional neural network architectures (AlexNet, GoogLeNet, ResNet-50, ShuffleNet, SqueezeNet, and MobileNet v2) were evaluated for their classification performance using three types of chlorophyll fluorescence images. Among these, ResNet50 outperformed the others, particularly when features from all image types were fused at the average pooling layer, yielding a peak classification accuracy of 98.61%. Uniform manifold approximation and projection analysis further validated the high discriminative power of the fused features. Experimental consistency was ensured through hydroponic growth of seedlings under controlled environmental conditions, with NaCl concentrations varied from 0 to 200 mM. Soybean seeds were surface-sterilised, germinated in the dark, and grown hydroponically for 7 days before exposure to salt. Chlorophyll fluorescence images and SPAD values were collected after 3 days of salt treatment, enabling accurate physiological assessments. A higher SPAD value generally indicates a higher chlorophyll concentration and thus potentially healthier plants. This study underscores the potential of combining fluorescence imaging with deep learning for precise salt stress diagnosis [84].

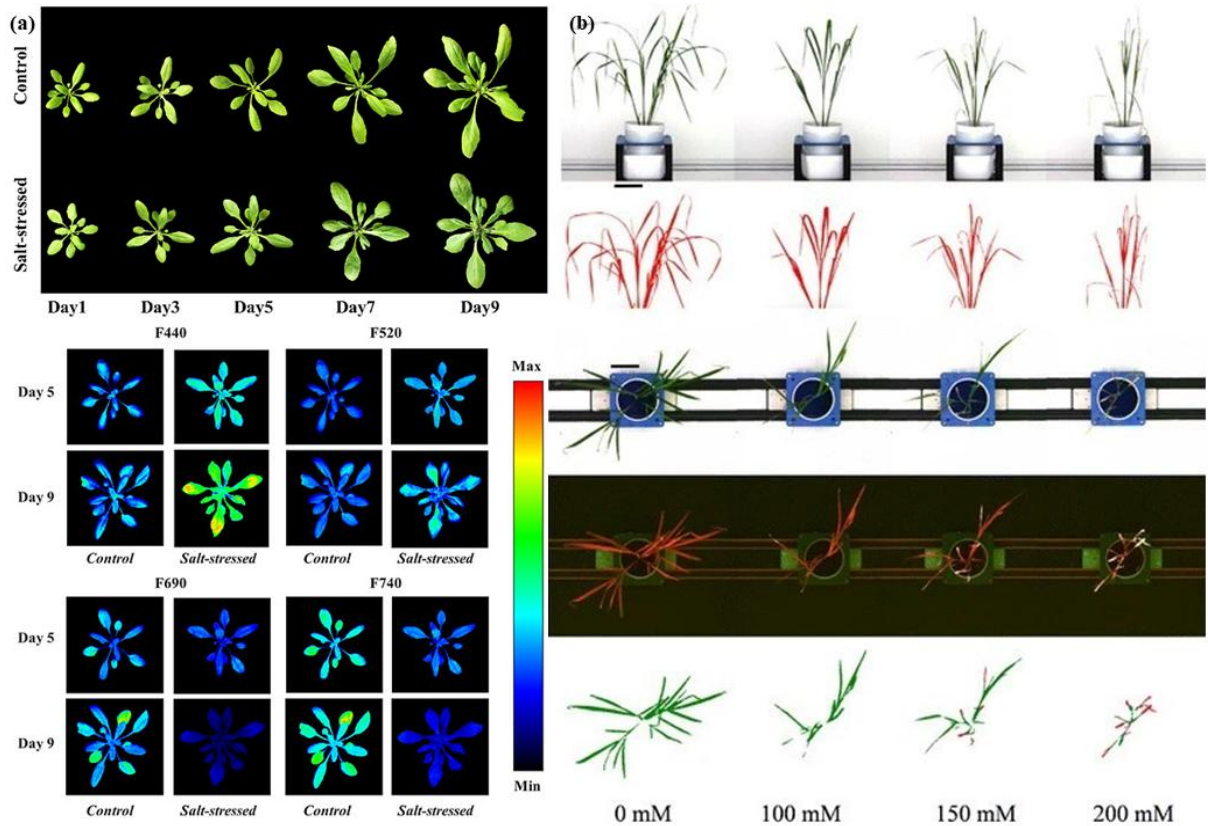

Figure 4 (a) RGB images of plants under control and salt-stress conditions at different time points. Pseudo-colour images depict fluorescence parameters F440, F520, F690, and F740 for plants under both treatments. Representative pseudo-colour images for F440 and F520 are shown for days 5 and 9, as well as for F690 and F740 on the same days post salt-stress treatment. The colour scale to the right of the images ranges from black (minimum value) to red (maximum value) [82]. (b) Salt stress was applied two weeks after transplantation. Side-view RGB images are presented, followed by object segmentation results used for size quantification. Top-view RGB images of the same plants are shown alongside corresponding fluorescence images captured from above. Colour-classified fluorescence images highlight healthy leaf tissue in green and senescent areas in purple [83].

## 4.2 Hyperspectral Imaging Applications

Salinity stress disrupts the physiological and biochemical processes in plants. A study was conducted in which four wheat lines were grown hydroponically under both control and salt stress conditions using 200 mM NaCl. Hyperspectral images were acquired one day after salt application, prior to the onset of any visible stress symptoms, enabling early detection. Successive volume maximisation identified two spectral endmembers representing control and salt-stressed conditions, which facilitated subsequent data simplification. The study introduced a vector-wise similarity measurement technique that transformed complex high-dimensional hyperspectral data into one-dimensional grayscale images while retaining critical spectral information. Two independent analytical methods—minimum difference of pair assignments and Bayesian classification—were applied to these grayscale images, producing consistent salt tolerance rankings. These rankings correlated strongly with conventional phenotypic assessments and historical tolerance data, demonstrating that HSI combined with innovative analytical techniques can objectively and quantitatively rank wheat genotypes early in stress onset, offering a powerful tool for breeding programs, as shown in Figure 5(b) [85].

Salinity stress heterogeneity in coastal shrub species *Myrica cerifera* and *Iva frutescens* was investigated across contrasting climatic conditions using physiological metrics and hyperspectral reflectance indices. Despite uniform relative water content and water band index (WBI), spatial variation in tissue chloride concentrations revealed uneven salt stress. The physiological reflectance index (PRI) effectively captured this heterogeneity and correlated strongly with photosynthetic efficiency ($R^2 = 0.79$ for *M. cerifera*, 0.72 for *I. frutescens*), underscoring its physiological sensitivity. Species-specific differences emerged, with stable PRI in *M. cerifera* and interannual PRI shifts in *I. frutescens*, likely due to flooding [86]. Further site-specific analysis of *M. cerifera* under varying sea spray exposure confirmed the robustness of PRI at both canopy and landscape scales, correlating with stomatal conductance, photosynthesis, and chlorophyll fluorescence ($R^2 = 0.69$), while NDVI and WBI showed limited sensitivity to salt stress. Airborne hyperspectral imagery validated the capacity of PRI to detect spatial stress variation, highlighting its value for early detection of salinity stress and its complementarity to NDVI in capturing distinct physiological responses in coastal vegetation [87].

The ratio of $Na^+$ to $K^+$ ions is pivotal in determining the extent of ionic toxicity and osmotic imbalance in plants subjected to salinity stress. Traditional ion quantification techniques involve destructive sampling and laborious chemical assays, unsuitable for high-throughput phenotyping. This study evaluated hyperspectral imaging as a rapid, non-destructive method for monitoring $Na^+$ and $K^+$ concentrations in a rice recombinant inbred line population derived from salt-sensitive and salt-tolerant parents. Hyperspectral data were collected under saline hydroponic conditions (EC = 9 dS m$^{-1}$), and spectral preprocessing pipelines were optimised for whole-plant imaging. Partial least squares regression models were developed using mean pixel spectra, smoothened spectra, and selected wavelength bands. Smoothened and filtered spectral datasets yielded superior prediction accuracies, confirming the feasibility and enhanced performance of hyperspectral imaging over conventional methods for real-time ionic monitoring, enabling large-scale screening of genetic variation in salinity tolerance [88].

Remote sensing of soil salinity through satellite-derived hyperspectral vegetation indices offers a non-invasive approach to monitor salinity-induced stress in agricultural fields. This study assessed 21 existing vegetation indices and developed novel indices that incorporate chlorophyll and water absorption wavelengths, using Hyperion hyperspectral imagery and soil salinity data from 108 sugarcane field locations. Vegetation indices integrating chlorophyll absorption bands or combined chlorophyll-water bands exhibited the strongest correlations with soil salinity, whereas indices relying solely on water absorption or visible bands were less predictive. The optimised soil-adjusted vegetation index exhibited a high training-phase correlation ($R^2 = 0.69$) but showed limited validation robustness. Newly developed salinity and water stress indices (SWSI-1, SWSI-2, SWSI-3) alongside the Vogelmann red edge index demonstrated improved validation performance, with root mean square errors of 1.14 to 1.17 dS/m, highlighting their promise for accurate soil salinity estimation in crop fields [89].

The impact of salinity on lettuce growth was investigated using four NaCl treatments (50, 100, 150 mM). After two weeks, leaf osmotic potential and water content were measured alongside hyperspectral imaging of 40 leaves. Preprocessing via Savitzky–Golay smoothing and standard normal variate normalisation produced 32,000 mean spectra. Principal component analysis on a calibration subset enabled development of an initial salinity detection model, complemented by an index approximating the second derivative in the red edge region. The application of both models to complete hyperspectral datasets generated artificial images that effectively discriminated between calibration and validation samples. Statistical analysis confirmed the ability of the model to detect salinity-induced physiological changes non-destructively, underscoring the utility of hyperspectral imaging for early salinity stress monitoring in leafy vegetables, as shown in Figure 5(a) [90].

A multi-year field study in the North China Plain assessed hyperspectral vegetation indices for monitoring soil salinity under brackish water irrigation, which often exacerbates soil salinity and reduces wheat yield. Spectral reflectance measurements during the grain-filling stage of winter wheat revealed that optimized vegetation indices incorporating salt-sensitive blue, red-edge, and near-IR bands correlated strongly with soil salinity at multiple soil depths, with the best performance at 30 cm depth ($R^2 = 0.81$, RMSE = 0.36 g/L). Linear and quadratic models based on normalized difference vegetation indices further provided robust soil salinity estimates ($R^2 \geq 0.63$, RMSE $\leq 0.62$ g/L) across all depths examined. These results demonstrate the practical applicability of hyperspectral remote sensing for precise and cost-effective soil salinity assessment, which is critical for irrigation management in crops [27].

Effects of drought and salinity stress on photosynthetic performance, leaf water content, and antioxidative enzyme activity were studied in three Greek olive cultivars using hyperspectral reflectance data collected with a spectroradiometer. Multivariate regression and classification analyses linked spectral signatures with physiological and biochemical parameters. Drought and salinity caused reductions in photosynthesis and leaf water content, while antioxidative enzyme activities increased, with the cultivar displaying higher stress tolerance. Normalised difference vegetation indices correlated strongly with photosynthetic rate, followed by PRI. Meanwhile, specific spectral regions were identified as indicators for each antioxidative enzyme. These pioneering findings demonstrate the potential for drone-mounted hyperspectral cameras to facilitate high-throughput phenotyping of plant stress responses, supporting precision agriculture in stress-prone environments [91].

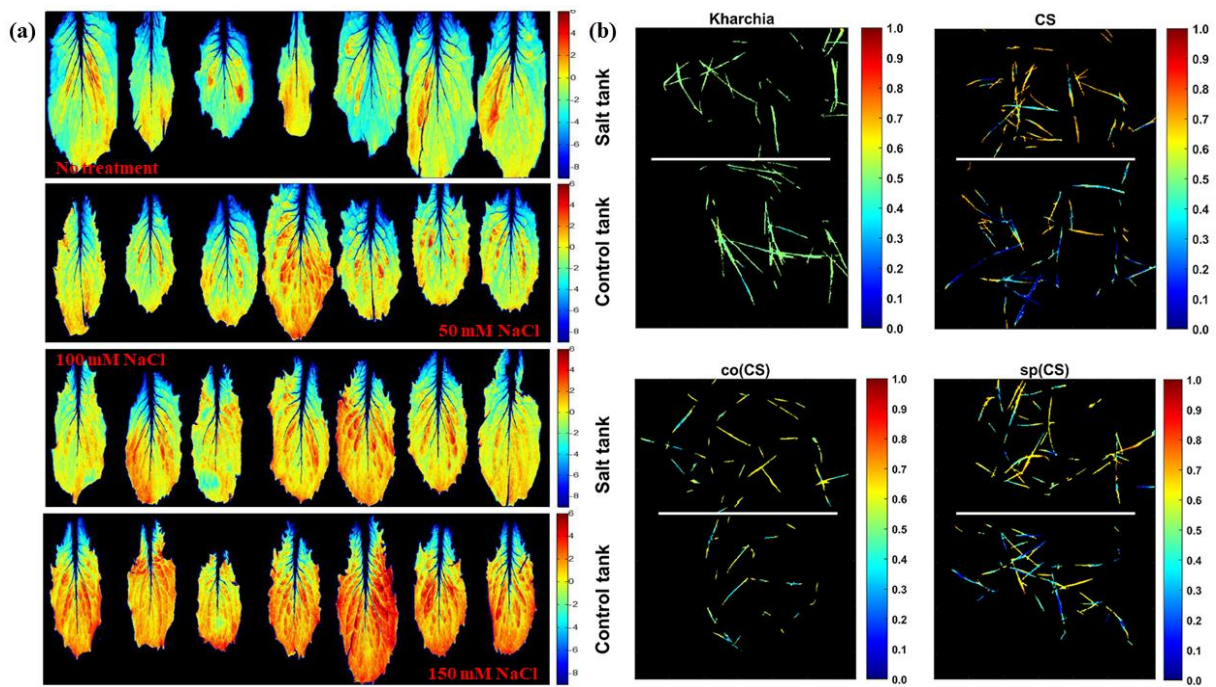

Figure 5 Virtual representations of scores obtained by projecting the hyperspectral images. (a) Leaves from the calibration set, where pixel colours range from blue to red, indicating increasing levels of salinity stress [90]. (b) Posterior probability of belonging to the salt-affected class based on similarity to the salt endmember, shown for both control and salt-treated pixels across each wheat line [85].

**4.3 Thermal Imaging Applications**

A non-destructive thermal imaging technique was employed to investigate the stomatal response of *Arabidopsis thaliana* subjected to salinity stress under high light conditions. Plants were exposed to varying concentrations of NaCl (75 mM, 150 mM, and 220 mM) to simulate short-term salt stress. Time-resolved thermograms revealed dynamic alterations in leaf temperature distribution, indicative of changes in stomatal behaviour under the dual stressors. Initial thermal responses, which closely mirrored changes in stomatal aperture, displayed exponential temperature kinetics. By applying a single-exponential model to the thermal data, time constants characterising thermal responses under acute light exposure were extracted. Salt-induced disruptions in stomatal functioning were evident by reduced stomatal conductance and transpiration rates, resulting in elevated rosette temperatures and shortened thermal time constants compared to control plants. Moreover, net carbon dioxide assimilation exhibited a concentration-dependent pattern, showing a decline at 220 mM NaCl but a modest increase at 75 mM. All salinity treatments under excessive light resulted in significant reductions in the maximal quantum yield of photosystem II, indicating photoinhibition exacerbated by salinity, as shown in Figure 6(a) and 6(b) [79].

Influence of salinity stress on alfalfa transpiration and growth was examined across four salinity levels: no salinity (S0), slight (S3), moderate (S5), and severe (S7). Canopy temperature data were collected using thermal IR remote sensing and analyzed using the three temperature (3T) model to estimate transpiration rates. The integration of imaging with the 3T model yielded highly accurate predictions, with coefficients of determination exceeding 0.8. A consistent increase in canopy temperature was observed with rising salinity levels, with temperature increments of 0.88°C, 0.98°C, and 1.19°C from S3 to S7, respectively. Alfalfa growth was markedly inhibited, as reflected in reductions in leaf area index by 19.3%, 31.2%, and 54.2%. Transpiration rates declined by 1.2%, 6.7%, and 20.0%, respectively, across increasing salinity levels. These changes were attributed to stunted plant growth and decreased canopy coverage under heightened salt stress [92].

A study on Nigella sativa (black cumin) evaluated the application of thermal imaging to assess salinity stress effects on leaf temperature. Plants were irrigated with water at four salinity levels—0.6, 1.5, 2.5, and 5.0 dS/m—prepared using various salt sources including $CaCl_2$, $MgCl_2$, NaCl, $Ca(NO_3)_2$, $MgSO_4$, and $Na_2SO_4$. Although the type of salt source did not significantly impact leaf temperature when averaged across salinity levels, a distinct trend of increasing leaf temperature with rising salinity was evident. Salinity impact on leaf temperature followed the sequence: 5.0 dS/m > 2.5 dS/m > 1.5 dS/m ≥ 0.6 dS/m. The highest temperatures were recorded at 5.0 dS/m for most salt sources; however, $MgSO_4$ and $Na_2SO_4$ did not significantly differ from the 2.5 dS/m treatment. A strong negative correlation between plant water consumption and leaf temperature was observed under CaCl$_2$ and MgSO$_4$ treatments, while a moderate negative correlation was seen for MgCl$_2$ and NaCl, suggesting a physiological link between reduced transpiration and elevated canopy temperature under salt stress conditions [93].

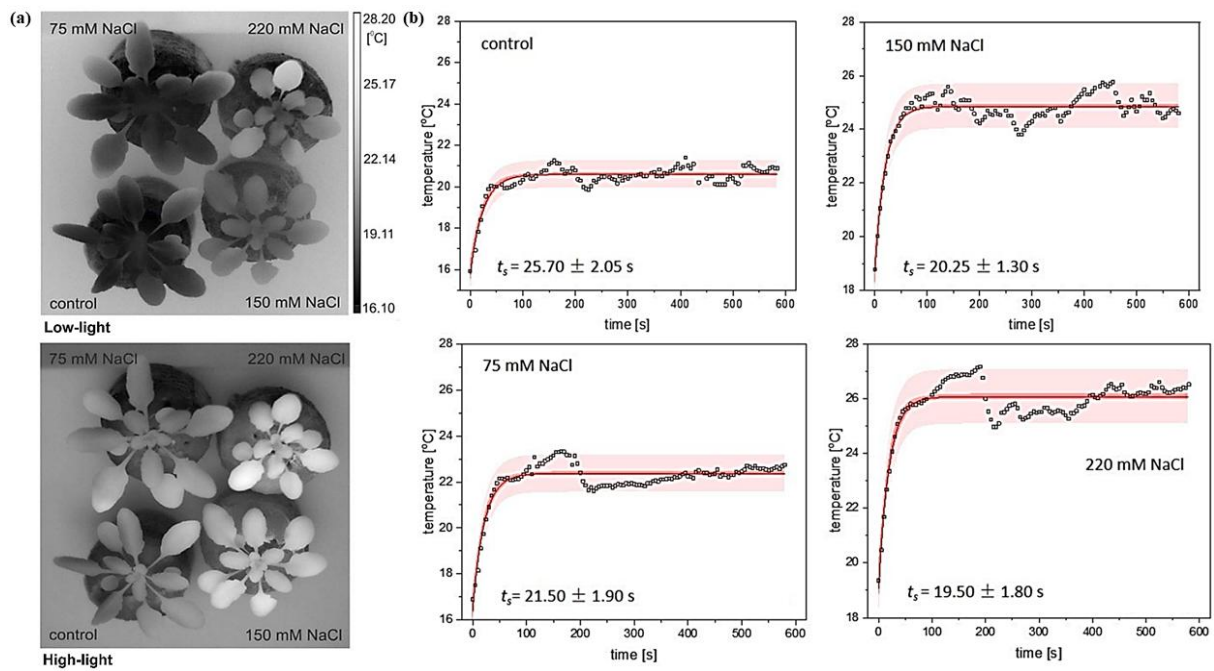

Figure 6 (a) IR thermograms of low-light-grown and high-light-treated *Arabidopsis thaliana* plants, acquired three days after salinity exposure, showing the average temperature distribution over the leaf rosettes of control and salt-treated plants. (b) Changes in rosette leaf temperature induced by acute high-light treatment in A. thaliana plants exposed to salinity. Each point represents the mean temperature from at least three randomly selected regions on thermal images for both control and NaCl-treated plants. Theoretical curves approximate the initial plant responses to excess light three days after salinity exposure [79].

A recent investigation explored the integration of thermal IR and RGB-derived indices with artificial neural networks to evaluate physiological and morphological traits in plants under saline conditions. Eighteen recombinant inbred lines and three parental genotypes were irrigated with saline water (150 mM NaCl), and assessed for relative water content, chlorophyll *a*, chlorophyll *b*, total chlorophyll, and plant dry weight. Substantial genetic variation was noted among genotypes for all traits. The normalized relative canopy temperature index demonstrated strong correlations with the measured traits, yielding R² values between 0.50 and 0.84. Several RGB-based indices, particularly the visible atmospherically resistant index, also correlated significantly with normalized relative canopy temperature, relative water content, chlorophylls, and plant dry weight, with R² values ranging from 0.49 to 0.62 across two seasons. Artificial neural network models further enhanced trait prediction, achieving R² values of 0.62–0.90 for training and 0.46–0.68 for cross-validation datasets. These results underscore the potential of integrating thermal-RGB imaging and machine learning for high-throughput phenotyping in saline environments [94].

Table 1 Comparison of imaging technologies for assessing salinity stress in plants.

| Parameter | Hyperspectral Imaging | Fluorescence Imaging | Thermal Imaging |
|---|---|---|---|
| Detection Mechanism | Detects changes in leaf pigment composition, water content, and biochemical alterations induced by salinity | Measures changes in chlorophyll fluorescence reflecting impaired photosynthesis due to salt stress | Detects altered leaf temperature caused by reduced transpiration under salt stress |
| Key Indicators | Spectral indices related to chlorophyll, carotenoids, water absorption bands, and stress pigments | Decreased variable fluorescence to maximum fluorescence and other chlorophyll fluorescence parameters indicate photosynthetic damage | Increased leaf temperature due to stomatal closure and reduced transpiration |
| Sensitivity | High sensitivity to biochemical and physiological changes induced by salt stress | Sensitive to early photosynthetic impairment before visible symptoms | Sensitive to physiological changes, but indirect for salinity effects |
| Temporal Resolution | Moderate, depends on scanning speed | High, real-time monitoring possible | High, suitable for continuous monitoring |
| Spatial Resolution | High, suitable for detailed mapping of stress variability | High, can detect heterogeneity in photosynthetic activity | Moderate to high, spatial patterns of temperature changes visible |
| Advantages | Enables early detection of subtle biochemical changes, spatial mapping of stress severity | Detects early photosynthetic impairment, non-destructive | Real-time, non-invasive measurement of transpiration-related stress |
| Limitations | Requires complex data processing; expensive equipment | Limited mainly to chlorophyll-related stress indicators | Environmental factors (ambient temperature, humidity) can affect readings |
| Typical Use Case Examples | Mapping salt-induced nutrient imbalances, pigment degradation, water content changes | Monitoring photosynthetic efficiency decline under salt stress | Assessing stomatal closure and drought-like symptoms caused by salinity |

Table 2 Summary of literature on the use of optical imaging technologies for assessing salinity stress in plants.

| Year | Plants | Salt Concentrations | Optical Technique | Data Analytic Technique | Excitation Range | Detection Range | Summary |
|---|---|---|---|---|---|---|---|
| 2008 [86] | Myricaceae and Asteraceae | 5 M NaNO$_3$ | Airborne Hyperspectral Imaging | Analysis of variance | Natural light condition | 384–1000 nm | Physiological Reflectance Index effectively detected salinity stress across the landscape. |
| 2008 [87] | *Myrica cerifera* | 5 M NaNO$_3$ | Airborne Hyperspectral Imaging | Frequency histograms | Natural light condition | 450-2450 nm | Chlorophyll fluorescence and hyperspectral reflectance were used to detect salinity stress in the evergreen coastal shrub. |
| 2013 [89] | Sugarcane | 1.9-9.2 dS/m | Airborne Hyperspectral Imaging | Correlation | Natural light condition | 427–2396 nm | Satellite-based hyperspectral remote sensing enables accurate soil salinity estimation in vegetated areas during the growing season. |
| 2014 [83] | Rice | 50 mM, 75 mM and 100 mM NaCl | Fluorescence Imaging | Nearest-neighbour colour classification | 400-500 nm | 500-750 nm | Salt concentrations had minimal impact on rice growth during the initial osmotic stress phase. |
| 2016 [90] | Lettuce | 50 mM, 100 mM and 150 mM NaCl | Hyperspectral Imaging | Principal Component Analysis | 400-1200 nm | 400–1000 nm | Marginal necrosis appearing in leaves of plants under salinity stress can be detected by the artificial images before the damage is visible. |
| 2018 [85] | Wheat | 200 mM NaCl | Hyperspectral Imaging | Bayesian method | Natural light condition | 400-900 nm | Two methods—minimum difference of pair assignments and a Bayesian approach—were used to analyze grayscale images, yielding similar rankings consistent with conventional phenotyping and historical salt tolerance data. |
| 2020 [92] | Alfalfa | NaCl, MgSO$_4$, and CaSO$_4$ (2:2:1) | Thermal Imaging | 3 T model | Natural light condition | 7.5–14 µm | Thermal infrared imaging was used to measure the canopy temperature of plants exposed to different salinity conditions. Transpiration rates were estimated using the three-temperature model. |
| 2021 [95] | Basil | 100 mM and 200 mM NaCl | Fluorescence and Multispectral Imaging | Principal Component Analysis | White (3000 K), 730 nm, 660 nm, 520 nm, 460 nm, and 405 nm | 400–1000 nm | Maximum fluorescence normalized difference vegetation index and leaf inclination are key traits distinguishing drought vs. non-stressed and drought vs. salinity-stressed plants. |
| 2021 [82] | *Arabidopsis* Columbia | 100 mM NaCl | Fluorescence Imaging | Principal Component Analysis and Support Vector Machine | 365 nm | 440 nm, 520 nm, 690 nm, and 740 nm | Classification accuracy at 5 days was 92.65%, increasing to 98.53% at 9 days under control and salt-stress treatments. |
| 2021 [27] | Wheat | 1 g/L TDS, 3 g/L TDS, and 5 g/L TDS | Hyperspectral Imaging | Non-Linear (Quadratic, Logarithmic, or Exponential) Models | Natural light condition | 378-1000 nm | Optimized spectral vegetation indices using salt-sensitive blue, red-edge, and near-infrared bands improved soil salinity retrieval ($R^2 \geq 0.58$, RMSE $\leq 0.62$ g/L), with highest accuracy at 30 cm depth ($R^2 = 0.81$, RMSE = 0.36 g/L). |
| 2021 [79] | Rosette | 75 mM, 150 mM, and 220 mM NaCl | Thermal Imaging | Time-dependent thermograms and box-and-whisker plots | White light | 7.5–14 µm | Decline in the maximal quantum yield of photosystem II under excessive salt level was observed in control and NaCl-treated plants |
| 2022 [91] | Olive | 50 mM NaCl | Hyperspectral Imaging | Principal Component Regression and Linear Discriminant Analysis | 400-1200 nm | 350-2500 nm | Correlation of hyperspectral imagery with photosynthetic rate and antioxidant enzyme activities established a basis for high-throughput plant phenotyping via drone-mounted hyperspectral cameras. |
| 2022 [88] | Rice | 270 mM NaCl with 9.9 mM CaCl$_2$ | Hyperspectral Imaging | Partial Least Squares Regression | Natural light condition | 550-1700 nm | Predictive capability of hyperspectral imaging-based method for monitoring Na$^+$ and K$^+$ levels in plants. |
| 2024 [84] | Soybean | 50 mM, 100 mM, 150 mM, and 200 mM NaCl | Fluorescence Imaging | Convolutional Neural Network | 650 nm | 650 nm | Achieved highest accuracy of 98.61% through fusion of features from chlorophyll fluorescence images. |
| 2024 [93] | Black Cumin | Salinity levels (1.5, 2.5, and 5.0 dS/m) for different salts: CaCl$_2$, MgCl$_2$, NaCl, Ca(NO$_3$)$_2$, MgSO$_4$, and Na$_2$SO$_4$. | Thermal Imaging | - | - | 7.5–13 µm | Effect of salinity levels on leaf temperature of black cumin. Leaf temperature increased with salinity, following the order: 5.0 dS/m > 2.5 dS/m > 1.5 dS/m ≥ 0.6 dS/m. The highest temperatures were observed at 5.0 dS/m with CaCl$_2$, MgCl$_2$, Ca(NO$_3$)$_2$, MgSO$_4$, and Na$_2$SO$_4$, though values for MgSO$_4$ and Na$_2$SO$_4$ were not significantly different from those at 2.5 dS/m. |
| 2024 [94] | Wheat | 150 mM NaCl | Thermal Imaging | Artificial Neural Networks | Natural light condition | 7.5–13 µm | Effectiveness of combining thermal infrared and RGB-derived indices with artificial neural networks to assess relative water content, chlorophyll a, chlorophyll b, total chlorophyll, and dry weight in 18 recombinant inbred lines and their three parents under saline irrigation (150 mM NaCl). |

## 5 Challenges and Limitations

Optical sensing technologies present a promising frontier for non-invasive monitoring of salinity stress in plants due to their capacity for rapid, large-scale data acquisition and real-time analysis. However, translating these technologies into real-world agricultural practices is fraught with substantial challenges and limitations. A key impediment is the non-specific nature of the visual and spectral signatures associated with salinity stress [5]. Phenotypic responses such as chlorosis, necrosis, wilting, and reductions in chlorophyll content are not exclusive to salt stress but are common among various abiotic stressors, including drought, heat, and nutrient imbalances. These overlapping symptoms hinder the diagnostic specificity of optical sensors, particularly in heterogeneous field conditions where multiple stressors often co-occur [6]. The absence of distinct spectral biomarkers uniquely indicative of salinity stress further compromises the diagnostic precision of conventional imaging modalities, such as multispectral and chlorophyll fluorescence sensors, thereby limiting their reliability under operational field conditions [96]. The issue is exacerbated by considerable inter- and intra-species heterogeneity in plant physiological responses to salinity. Various species and genotypes utilize distinct adaptation strategies—such as ion compartmentalization, selective absorption or exclusion of sodium and chloride ions, and osmotic adjustment—that variably influence optical characteristics. These physiological changes result in diverse spectral and reflectance signatures, hindering the creation of universal detection algorithms. For example, younger leaves may display stress-induced optical alterations sooner than older leaves owing to variations in cuticle thickness, stomatal activity, and chloroplast density [43]. Consequently, spectral data obtained from heterogeneous canopies may not accurately reflect the overall stress condition of the crop. The genotype- and age-dependent optical variability requires the calibration of crop-specific and stage-specific sensing algorithms, hence constraining the scalability and generalizability of optical stress detection systems in diverse agricultural settings [97].

Environmental and operational factors add additional complexity to field-based optical sensors. Variations in ambient light circumstances, such as sun angle, cloud cover, and spectral composition, can substantially influence sensor readings. Furthermore, external disturbances from soil reflectance, neighboring vegetation, and surface water droplets can provide background noise that undermines signal integrity [19]. Physical dynamics, such as leaf fluttering caused by wind, can lead to motion artifacts or misalignment of sensor data, especially during airborne or mobile scanning. These issues require effective preprocessing methods, such as radiometric correction, normalization, and motion compensation, which are challenging to standardize in extensive deployments. The necessity for regulated environmental conditions to achieve dependable readings highlights a significant constraint in the application of optical sensors for field-scale, real-time monitoring [98].

A critical temporal constraint in optical detection of salinity stress lies in the latency between root-level salt accumulation and the onset of detectable foliar symptoms. Salinity stress often begins with ionic toxicity and osmotic imbalance in the root zone, long before visual cues emerge on above-ground tissues. This lag limits the utility of many optical systems in early detection, reducing their effectiveness as proactive diagnostic tools [40].

By the time visible changes such as chlorosis or leaf curling are observed, irreversible physiological damage may have occurred, diminishing the opportunity for timely agronomic intervention. Additionally, spatial limitations of certain platforms—especially unmanned aerial vehicles with lower spatial resolution—can hinder the detection of early or localized symptoms, such as those confined to leaf undersides or shaded canopy regions, where salt stress may initially manifest [97].

Advanced imaging techniques such as hyperspectral imaging, infrared cameras, and light detection and ranging systems provide superior capabilities by acquiring high-dimensional, spatially-resolved datasets. These technologies can identify nuanced biochemical and structural alterations linked to salinity stress, including variations in pigment content, cell turgor, and leaf shape. Nevertheless, the abundance of these databases incurs a significant computational burden [99]. High-dimensional data requires meticulous preparation, including noise filtration, spectrum calibration, and dimensionality reduction, alongside the implementation of advanced machine learning techniques to discern significant patterns. Developing dependable predictive models necessitates comprehensive training with well labeled information, often incorporating invasive ground-truth measures like leaf or root ion concentrations. The data collection techniques are laborious and expensive, rendering large-scale deployment unfeasible. Moreover, machine learning models frequently exhibit sensitivity to their training settings, resulting in restricted transferability to different crops, habitats, or seasons, hence diminishing their resilience in varied agricultural contexts [12], [13].

Economic and logistical constraints significantly hinder the widespread adoption of optical sensing technologies for salinity monitoring. High-performance instruments such as hyperspectral cameras and light detection and ranging units are prohibitively expensive for most farmers and require specialized expertise for installation, calibration, and data interpretation [61], [100]. In contrast, low-cost alternatives—including smartphone-based systems or consumer-grade multispectral sensors—lack the spectral sensitivity and resolution necessary for detecting subtle or early-stage salinity stress. This disparity between advanced research tools and field-ready technologies creates a substantial translational gap. Bridging this divide requires not only technological innovation to develop cost-effective, user-friendly platforms but also capacity-building initiatives to train users in their effective deployment and data utilization [16].

# 6 Conclusion

Salinity stress poses a significant risk to global agricultural sustainability, necessitating the immediate development of new, rapid, and scalable detection methods to monitor plant health in saline environments accurately. This review systematically evaluates the use of advanced optical spectroscopic imaging technologies—specifically fluorescence imaging, hyperspectral imaging, and thermal infrared imaging—as effective, non-destructive methods for detecting dynamic physiological and biochemical changes in plants subjected to salt stress. These imaging techniques offer exceptional benefits in terms of spatial and temporal resolution, enhancing high-throughput phenotyping initiatives and facilitating the selection of salt-tolerant genotypes in breeding programs. The implementation of these technologies on terrestrial platforms and their integration with aerial systems, including unmanned aerial vehicles and satellite sensors, substantially expands the monitoring capabilities, facilitating real-time, multi-scale evaluations of salinity effects across diverse agricultural terrains. These integrated technologies enable the creation of comprehensive salinity distribution maps, which are crucial for implementing site-specific, precision agricultural methods that enhance resource utilisation and crop yield.

Notwithstanding their potential, some technical challenges persist, including fluctuations in ambient conditions, crop-specific spectral fingerprints, and the absence of universally applicable calibration models. Future research should prioritise the development of standardised imaging techniques and robust machine learning frameworks to decode complex spectral data across diverse agronomic scenarios. The integration of optical imaging with complementary sensing technologies, such as soil electrochemical sensors and ion-selective probes, could significantly improve early detection and diagnostic precision. Progress in cost-effective, portable imaging devices will be crucial in disseminating these technologies from research institutions to agricultural practices, thereby revolutionising salinity stress management and enhancing the resilience of agricultural systems in salinising conditions.

**Statements and Declarations:**

**Competing Interests:** Authors declare that they have no competing interests.

**Funding:** The project is funded by the Parul University (Grant No. RDC/IMSL/207).